\documentstyle[11pt,IAUS215,twoside,epsf]{article}

\def\edcomment#1{\iffalse\marginpar{\raggedright\sl#1\/}\else\relax\fi}
\reversemarginpar

\begin{document}
\vspace*{1cm}
\title{The Rotational Velocity of Helium-rich Pre-White Dwarfs}
 \author{Thomas Rauch}
  \affil{Dr.-Remeis-Sternwarte, D-96049 Bamberg, Germany\\
         Institut f\"ur Astronomie und Astrophysik, D-72076 T\"ubingen, Germany}
 \author{Sebastian K\"oper, Stefan Dreizler, Klaus Werner}
  \affil{Institut f\"ur Astronomie und Astrophysik, D-72076 T\"ubingen, Germany}
 \author{Ulrich Heber}
  \affil{Dr.-Remeis-Sternwarte, D-96049 Bamberg, Germany}
 \author{I\@. Neill Reid}
  \affil{STScI, Baltimore, MD, USA}
  
\begin{abstract}
Previous investigations on hydrogen-rich white dwarfs generally yield only very small rotational velocities
($v_{\mathrm{rot}}\cdot\sin i$). 
We have analyzed line profiles in high-resolution optical spectra of eight 
hydrogen-deficient (pre-) white dwarfs and find deviations from the
dominant Stark line broadening in five cases which, interpreted as an
effect of stellar rotation, indicate projected 
rotational velocities of $40 - 70 \mathrm{km/sec}$. For the three least luminous 
stars upper limits of $v_{\mathrm{rot}}\cdot\sin i = 15 - 25 \mathrm{km/sec}$ could be derived only.
The resulting velocities correlate with luminosity and mass. However, since the mass-loss rate is
correlated to the luminosity of a star, the observed line profiles may be
affected by a stellar wind as well. In the case of RX\,J2117.1+3412\index{RX J2117.1+3412}, this
would solve discrepancies to results of pulsational modeling ($v_{\mathrm{rot}}\cdot\sin i \approx 0$).  
\end{abstract}

Detailed analyses of the NLTE Balmer line cores of DA white dwarfs have
revealed that for the vast majority the rotational velocities are extremely
low, so that they cannot be detected, even with high-resolution high-S/N
spectra (Heber et al\@. 1997, Koester et al\@. 1998). 
This unexpected behavior poses the question how these WDs have lost
their angular momentum during previous evolutionary phases. To our best
knowledge no one has ever tried a similar analysis on non-DA white dwarfs or
hydrogen-deficient pre-white dwarfs. The only exception is the PG\,1159-type
central star of NGC\,246\index{NGC 246}, whose high rotational rate is well known since many
years (Heap 1975, Rauch \& Werner 1997). High-resolution optical (HIRES at
KECK, EMMI at ESO NTT) spectra of PG\,1159
stars can in principle be used to resolve the narrow line cores (in absorption
or emission) of He\,{\sc ii}, C\,{\sc iv}, and O\,{\sc vi} and to constrain the
rotational velocity.

Synthetic spectra of NLTE model atmospheres (see Werner et al\@. 1997 for parameters and references) 
have been convolved with rotational profiles in order to compare them to the observation. 
Different lines have been evaluated for the determination of the rotational velocity. 
The adopted $v_{\mathrm{rot}}\cdot\sin i$
are summarized in Fig\@. 1. However, we are aware of some problems (K\"oper et al\@. 2001). 
\vspace{3mm}

\noindent
\begin{minipage}[t]{6cm}
\setlength{\textwidth}{6cm}
\setlength{\textheight}{6cm}
\setlength{\oddsidemargin}{-3cm}
\addtolength{\topmargin}{-3cm}
\epsfxsize=6cm
\epsffile{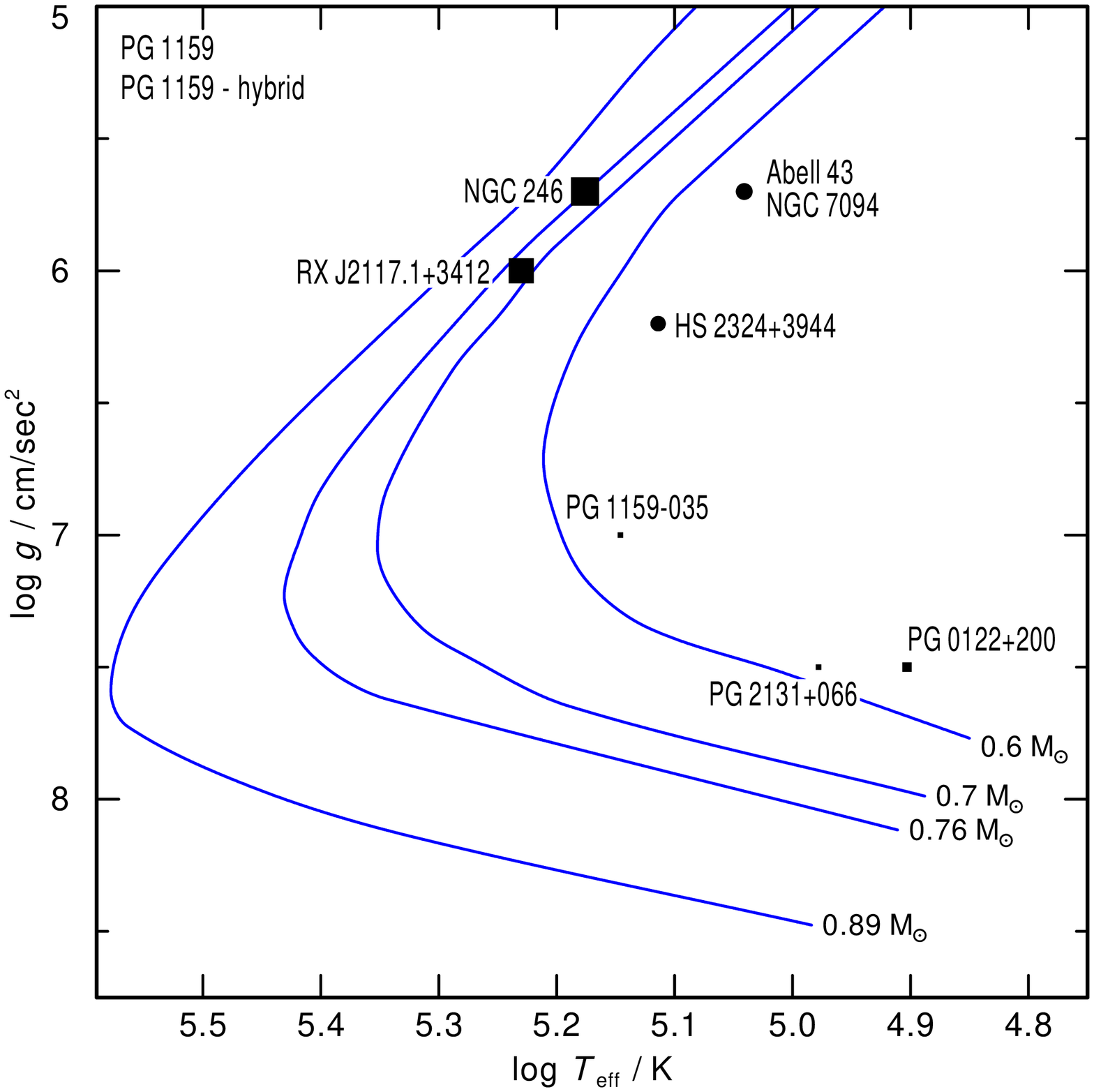}
\end{minipage}
\vspace{-6.2cm}

\noindent
\hbox{}\hspace{6.2cm}
\begin{minipage}[t]{7.0cm}
\setlength{\textwidth}{7.0cm}
\setlength{\textheight}{6cm}
\setlength{\oddsidemargin}{-3cm}
\addtolength{\topmargin}{-3cm}
\small
Fig.\,1. Position of our programme stars in the $\log T_{\mathrm{eff}} - \log g$ plane
compared to evolutionary tracks of helium-burning post-AGB stars (Wood \& Faulkner 1986, labeled by  
the stellar mass). The symbol's size is equivalent to $v_{\mathrm{rot}}\cdot\sin i$ of the star:
\vspace{1.5mm}

\begin{tabular}{lrll}
PG\,1159$-$035\index{PG 1159-035}       &  $< 15$ &                & km/sec \\
PG\,2131+066\index{PG 2131+066}         &  $< 15$ &                & km/sec \\
PG\,0122+200\index{PG 0122+200}         &  $< 25$ &                & km/sec \\
Abell\,43\index{Abell 43}               &  $  42$ & \raisebox{-0.7mm}{$^{\pm 13}$}       & km/sec \\
HS\,2324+3944\index{HS 2324+3944}       &  $  42$ & $^{+28}_{-12}$ & km/sec \\       
NGC\,7094\index{NGC 7094}               &  $  46$ & \raisebox{-0.7mm}{$^{\pm 16}$}       & km/sec \\
RX\,J2117.1+3412\index{RX J2117.1+3412} &  $  68$ & $^{+14}_{-18}$ & km/sec \\       
NGC\,246\index{NGC 246}                 &  $  77$ & $^{+23}_{-17}$ & km/sec \\       
\end{tabular}

\end{minipage}
\vspace{3mm}

All five low-gravity (i.e.\,high-luminosity) PG\,1159 stars show deviations
from synthetic ``Stark-broadened'' line profiles in the observed line cores.
This suggests that the more massive of these have higher rotational velocities (Fig.\,1).
However, the rapid rotation of 
RX\,J2117.1+3412\index{RX J2117.1+3412} ($v_{\mathrm{rot}} \sin i = 68 \mathrm{km/sec}$) is in contradiction
to the asteroseismology results of Vauclair et al\@. (2002). They found $v_{\mathrm{rot}} \sin i < 0.5 \mathrm{km/sec}$.
Although most of their parameters derived from the pulsational modeling, 
e.g\@. $T_{\mathrm{eff}}$, mass, distance, and luminosity, are in 
disagreement with results from spectral analysis (Rauch \& Werner 1997),
this may be a hint for an influence of the stellar wind on the observed line
profiles, i.e\@. the measured line profiles may be affected by an
additional turbulence broadening in the wind of this relatively luminous star.
The fast ``rotation'' of NGC\,246\index{NGC 246} may then be explained by wind effects as well.
The slow rotation of PG\,1159-035, deduced from asteroseismology (Winget et al\@. 1991), is corroborated.
Silvotti et al\@. (1999) presented an asteroseismology study of HS\,2324+3944\index{HS 2324+3944} but they could not
achieve sufficient precision in their results.\vspace{1mm}
\newline
This research was supported by the DLR (grants 50\,OR\,9705\,5 and 50\,OR\,0201).

\end{document}